# Voltage-controlled long-range propagation of indirect excitons in van der Waals heterostructure

L. H. Fowler-Gerace,[1] D. J. Choksy,[1] and L. V. Butov[1]

[1]*Department of Physics, University of California at San Diego, La Jolla, CA 92093, USA*

(Dated: January 4, 2021)

Indirect excitons (IXs), also known as interlayer excitons, can form the medium for excitonic devices whose operation is based on controlled propagation of excitons. A proof of principle for excitonic devices was demonstrated in GaAs heterostructures where the operation of excitonic devices is limited to low temperatures. IXs in van der Waals transition-metal dichalcogenide (TMD) heterostructures are characterized by high binding energies making IXs robust at room temperature and offering an opportunity to create excitonic devices operating at high temperatures suitable for applications. However, a characteristic feature of TMD heterostructures is the presence of moiré superlattice potentials, which are predicted to cause modulations of IX energy reaching tens of meV. These in-plane energy landscapes can lead to IX localization, making IX propagation fundamentally different in TMD and GaAs heterostructures and making uncertain if long-range IX propagation can be realized in TMD heterostructures. In this work, we realize long-range IX propagation with the $1/e$ IX luminescence decay distances reaching 13 microns in a $MoSe_2/WSe_2$ heterostructure. We trace the IX luminescence along the IX propagation path. We also realize control of the long-range IX propagation: the IX luminescence signal in the drain of an excitonic transistor is controlled within 40 times by gate voltage. These data show that the long-range IX propagation is possible in TMD heterostructures with the predicted moiré superlattice potentials.

PACS numbers:

Spatially indirect excitons (IXs) are formed by electrons and holes confined in separated layers. The separation between the electron and hole layers allows for controlling the overlap of electron and hole wave functions and achieving long IX lifetimes, orders of magnitude longer than lifetimes of spatially direct excitons (DXs) [1]. The long IX lifetimes allow them to travel over long distances before recombination [2–11].

IXs have built-in dipole moments $ed$ ($d$ is the separation between the electron and hole layers) and their energy can be controlled by voltage: Gate voltage $V_g$ controls the electric field normal to the layers $F_z \propto V_g$ and changes the IX energy by $edF_z$. This allows for creating tailored in-plane potential landscapes for IXs $E(x,y) = -edF_z(x,y)$ and controlling them in situ by voltage $V_g(x,y)$. The possibility to control IX energy by voltage and the ability of IXs to propagate over long distances led to the realization of a variety of tailored voltage-controlled in-plane potential landscapes, which are explored in studies of IX transport. These landscapes include excitonic ramps [2, 6], excitonic lattices [12–14], excitonic narrow channels [15, 16], excitonic conveyers [17], and excitonic split gate devices [18].

IX devices are also explored for developing signal processing based on the exciton dipole, that is a novel computational state variable, different from established computational state variables such as electron charge in electronic devices. Potential advantages of excitonic devices include energy-efficient signal processing and seamless coupling to optical communication [11]. Experimental proof-of-principle was demonstrated for excitonic transistors [19, 20].

The realization of excitonic devices, whose operation is based on controlled propagation of excitons, relies on meeting the requirements of (i) long-range IX propagation over lengths exceeding the in-plane dimensions of excitonic devices and (ii) in situ control of IX energy and IX propagation, in particular, by voltage. These requirement were met with IXs in GaAs heterostructures and the studies outlined above used the GaAs platform.

However, excitons exist at temperatures roughly below $E_{ex}/k_B$ ($E_{ex}$ the exciton binding energy, $k_B$ the Boltzmann constant) [21] and, due to their low binding energies, IXs in GaAs heterostructures exist at low temperatures. $E_{ex}$ is typically $\sim 4$ meV in GaAs/AlGaAs heterostructures [22] and achieves $\sim 10$ meV in GaAs/AlAs heterostructures [23]. The proof of principle for the operation of IX switching devices based on voltage-controlled IX propagation was demonstrated up to $\sim 100$ K in GaAs heterostructures [20]. IX devices based on controlled IX propagation are also explored in GaN/AlGaN heterostructures with high $E_{ex}$ reaching $\sim 30$ meV [24].

Van der Waals heterostructures composed of atomically thin layers of TMD [25] allow the realization of excitons with high binding energies [26, 27]. IXs in TMD heterostructures are characterized by binding energies reaching hundreds of meV [28, 29] making them stable at room temperature [30]. Due to the high IX binding energy, TMD heterostructures can form a material platform for creating excitonic devices operating at high temperatures suitable for applications.

However, in contrast to GaAs heterostructures, a characteristic feature of TMD heterostructures is the presence of significant moiré superlattice potentials, which are predicted to cause modulations of IX energy reaching tens of meV [31–34]. These in-plane energy landscapes can lead to IX localization, making IX propagation fundamentally different in TMD and GaAs heterostructures and making uncertain if long-range IX propagation can be realized in TMD heterostructures with moiré superlattice potentials.

Propagation of both DXs in TMD monolayers [35–40] and IXs in TMD heterostructures [41–46] is intensively studied. A relatively short-range IX propagation with $1/e$ IX luminescence decay distances reaching $\sim 3$ $\mu$m [41–45], control of IX propagation by voltage within these distances [43, 44], and control of DX luminescence by voltage up to 5 $\mu$m away from the generation spot [46] were reported in TMD heterostructures.

The extent to which the moiré superlattice potentials affect the IX diffusivity is not fully established. The comparison of MoSe$_2$/WSe$_2$ heterostructures with MoSe$_2$/hBN/WSe$_2$ heterostructures where the moiré superlattice potential is suppressed by an hBN spacer between the MoSe$_2$ and WSe$_2$ layers evidences the reduction of IX propagation due to moiré superlattice potentials: The IX propagation with the $1/e$ decay distance beyond the laser spot in the range of fractions of $\mu$m was observed in MoSe$_2$/WSe$_2$, and the IX propagation with the $1/e$ decay distance up to 2.6 $\mu$m was observed in MoSe$_2$/hBN/WSe$_2$ [43].

In this work, we realize in a MoSe$_2$/WSe$_2$ TMD heterostructure IX propagation with the $1/e$ IX luminescence decay distances reaching 13 $\mu$m, trace the IX luminescence along the IX propagation path, and control the IX propagation by voltage, presenting the direct measurement of voltage-controlled long-range IX propagation.

**Results**. The MoSe$_2$/WSe$_2$ heterostructure is assembled by stacking mechanically exfoliated 2D crystals on a graphite substrate. The MoSe$_2$ and WSe$_2$ monolayers are encapsulated by hexagonal boron nitride (hBN) serving as dielectric cladding layers. The energy-band diagram is schematically shown in Fig. 1a. IXs are formed from electrons and holes confined in adjacent MoSe$_2$ and WSe$_2$ monolayers, respectively. The lowest energy DX state is optically active in MoSe$_2$ and dark in WSe$_2$, and the lowest energy IX state is optically active [31, 32, 47–51]. The bias across the heterostructure is created by the gate voltage $V_g$ applied between the narrow graphene top gate (Fig. 1b) and the global graphite back gate.

The graphene gate is centered at $x = 4$ $\mu$m (Fig. 1d-f). The heterostructure region to the left of the gate ($x = -7$ to 2 $\mu$m) is referred to as the source and the heterostructure region to the right of the gate ($x = 6$ to 13 $\mu$m) is referred to as the drain. IXs are optically generated by laser excitation focused in the source region.

When the device is in the off state, IX propagation from the source to the drain is suppressed and the IX luminescence profile follows the laser excitation profile (Fig. 1e). When the device is switched on, IXs spread out away from the laser excitation spot and propagate to the drain region (Fig. 1d).

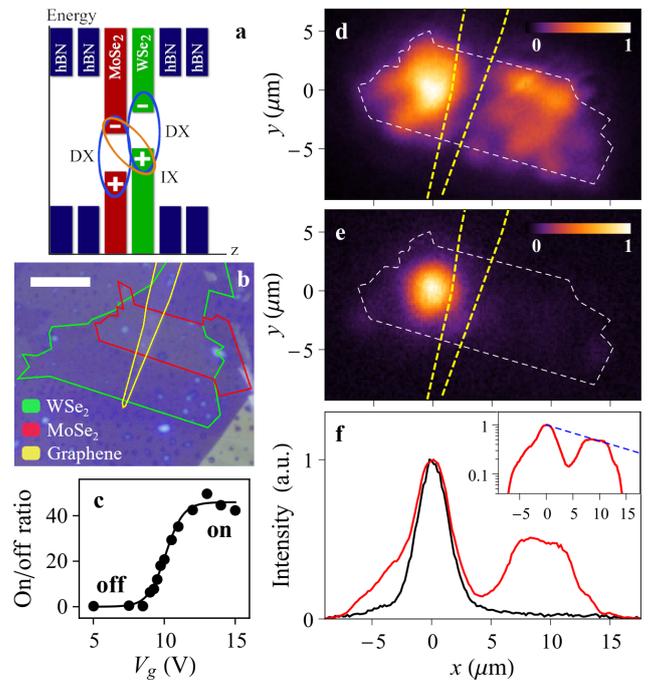

FIG. 1: **Voltage-controlled IX propagation.** (a) Band diagram of van der Waals MoSe$_2$/WSe$_2$ heterostructure. The ovals indicate a direct exciton (DX) and an indirect exciton (IX) composed of an electron (−) and a hole (+). (b) Microscope image showing the layer pattern of the device, scale bar is 10 $\mu$m. The green, red, and yellow lines indicate the boundaries of the WSe$_2$ and MoSe$_2$ monolayers and graphene gate, respectively. (d,e) $x - y$ images of IX luminescence in the on (d) and off (e) state of the excitonic transistor. The white and yellow dashed lines show the boundary of the MoSe$_2$/WSe$_2$ heterostructure and graphene gate, respectively. The gate voltage $V_g$ controls the IX propagation from the laser excitation spot at $x = 0$ (the source) to the other side of the graphene gate $x \gtrsim 6$ $\mu$m (the drain). $V_g = 10$ V (d), 0 (e). (f) Normalized IX luminescence profiles along $y = 0$ for the images in (d) and (e) shown by the red and black lines, respectively. Inset shows the same IX luminescence profile in the on state on log scale. For comparison, dashed line shows exponential signal reduction with $1/e$ decay distance 13 $\mu$m. A lower IX luminescence intensity is seen in the region covered by the graphene gate, which is centered at $x = 4$ $\mu$m. (c) Total IX luminescence intensity in the drain (integrated over $x = 6 - 13$ $\mu$m) vs $V_g$. For all data, $P_{ex} = 4$ mW, $T = 1.7$ K.

Figure 1f shows IX luminescence profiles along $y = 0$ for the images in Fig. 1d,e. Inset shows the same IX luminescence profile for the on state on log scale. For comparison, dashed line shows exponential signal reduction with the $1/e$ decay distance 13 $\mu$m. A lower IX luminescence intensity is seen in the region covered by the graphene gate. The total IX luminescence intensity in the drain region (integrated over $x = 6 - 13$ $\mu$m) increases by 40 times as the gate voltage switches from off to on (Fig. 1c). Figures 1c-f present voltage-controlled long-range propagation of IXs in MoSe$_2$/WSe$_2$ heterostructure.

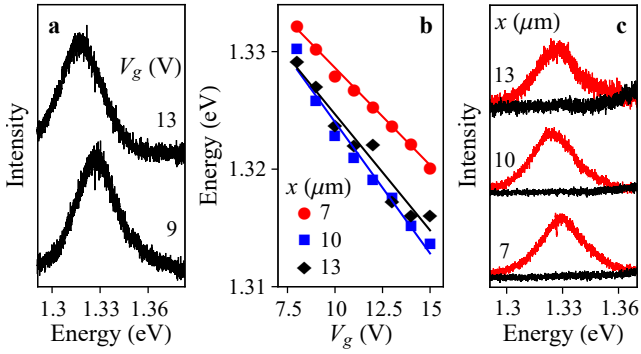

FIG. 2: **Voltage-controlled IX energy. Tracing IX luminescence along IX propagation.** (a) IX luminescence spectra at $x = 13$ μm for gate voltages $V_g = 9$ and 13 V. (b) IX energy vs $V_g$ for positions in the drain region $x = 7$, 10, and 13 μm. (c) IX spectra in on (red) and off (black) state of the excitonic transistor for positions in the drain region $x = 7$, 10, and 13 μm. $V_g = 10$ V (on), 8.5 V (off). For all data, $P_{ex} = 4$ mW, $T = 1.7$ K.

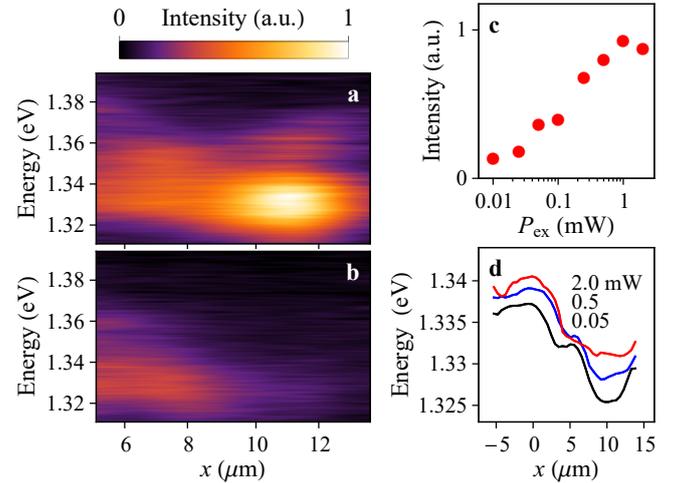

FIG. 3: **Excitation power dependence of IX propagation.** (a,b) $x$–energy images of IX luminescence in the drain region for $P_{ex} = 2$ mW (a) and 0.05 mW (b). (c) Total IX luminescence intensity in the drain (integrated over $x = 6 - 13$ μm) vs $P_{ex}$. (d) IX energy vs position for $P_{ex} = 0.05$ (black), 0.5 (blue), and 2 (red) mW. For all data, $V_g = 10$ V, $T = 1.7$ K.

Next, we verify if the propagating luminescence signal corresponds to the IX luminescence. Figures 2a,b show that at voltages, which allow for IX propagation across the sample, the exciton luminescence energy is controlled by voltage in the entire drain region. This exhibits the basic IX property – IX energy control by voltage, outlined in the introduction. Figure 2c presents tracing the IX luminescence along the IX propagation path all the way up to the heterostructure edge ca. 13 μm away from the region of IX optical generation. Figure 2c also shows that IX luminescence spectra at the drain demonstrate effective switching behavior between the off (black spectra) and on (red spectra) state. Tracing the IX luminescence along the IX propagation path with $1/e$ IX luminescence decay distance 13 μm and controlling IX propagation by voltage present the direct measurement of long-range IX propagation and excitonic transistor action in TMD heterostructures.

The IX luminescence spectra are traced over the drain region at different excitation powers $P_{ex}$ (Fig. 3a,b). The total IX luminescence intensity in the drain increases with $P_{ex}$ (Fig. 3c). Figures 3a-c show that the IX propagation enhances with excitation power.

Figure 3d shows the optically measured IX energy along the IX propagation path. The overall IX energy reduction is observed (i) with increasing separation from the IX optical generation spot and (ii) with reducing $P_{ex}$. Both indicate an increase of IX energy with increasing IX density. This is consistent with the repulsive interaction between IXs, which are dipoles oriented normal to the layers. Similar reduction of IX energy with increasing separation from the optical generation spot or with reducing $P_{ex}$ is also characteristic of IXs in GaAs heterostructures and is explained in terms of the repulsive interaction between IX dipoles [5]. The local IX energy variations in the range of a few meV (Fig. 3d) are likely caused by the lateral potential landscape across the heterostructure. The IX energy variations due to the moiré superlattice have the period in the nm range [52], these short-range energy variations are not resolved in the optical experiment with resolution 1.5 μm.

Figure 4 presents the temperature dependence of IX propagation. The long-range IX propagation through the drain region and the switching between on and off state are observed up to approximately 50 K.

**Discussion.** The phenomenological properties of voltage-controlled long-range IX propagation, outlined above, are different from those in GaAs heterostructures [19, 20]. GaAs/AlAs heterostructures, where IXs are formed from electrons and holes confined in adjacent AlAs and GaAs layers, respectively, have a staggered band alignment [20, 23] similar to MoSe$_2$/WSe$_2$ heterostructures (Fig. 1a). This makes GaAs/AlAs heterostructures a more close system for the comparison with MoSe$_2$/WSe$_2$ heterostructures. For the excitation spot positioned in the source region, similar to the experimental geometry in Fig. 1, in the GaAs/AlAs heterostructures IX propagation is long-range at $V_g = 0$ while both a positive and negative voltage on a gate creating a barrier and a trap for IXs, respectively, suppress the IX propagation [20]. This is clearly different from the properties of voltage-controlled IX propagation in the MoSe$_2$/WSe$_2$ heterostructure where applied voltage strongly enhances the IX propagation (Fig. 1).

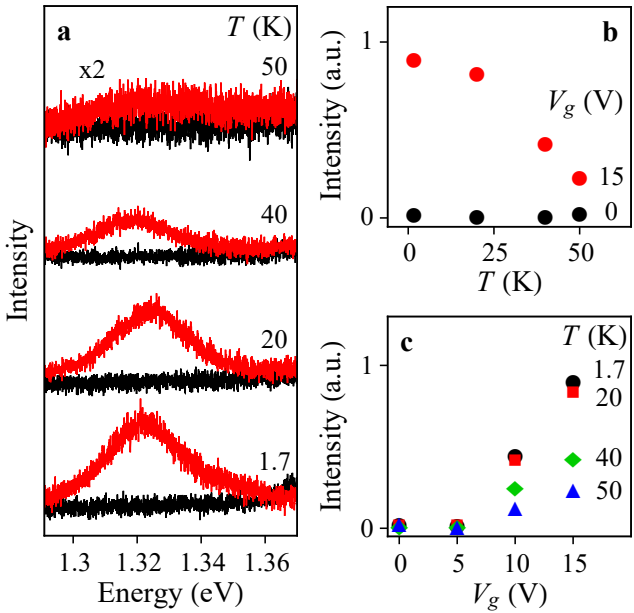

FIG. 4: **Temperature dependence.** (a) IX luminescence spectra in the drain at $x = 13$ μm at temperatures $T = 1.7, 20, 40$, and $50$ K in on (red) and off (black) state of the excitonic transistor. $V_g = 15$ V (on) and 0 (off). (b) Total IX luminescence intensity in the drain (integrated over $x = 6 - 13$ μm) vs temperature in on ($V_g = 15$ V) and off ($V_g = 0$) state. (c) Total IX luminescence intensity in the drain (integrated over $x = 6 - 13$ μm) vs $V_g$ for temperatures $T = 1.7, 20, 40$, and $50$ K. For all data, $P_{ex} = 4$ mW.

A different behavior is also observed for the excitation spot positioned on the gate electrode. For the GaAs/AlAs heterostructures, trapping (or anti-trapping) IX potentials created by gate voltage cause the IX cloud confinement in (or spreading away from) the gate region [20]. For the MoSe$_2$/WSe$_2$ heterostructure, the IX energy reduces with increasing $V_g$ that facilitates trapping IXs in the gate region. However, similar to the case of excitation in the source region (Fig. 1), an enhancement of the IX propagation away from the excitation spot with increasing $V_g$ is observed in the case of excitation in the gate region [Fig. S1 in supplementary information (SI)].

These distinctive differences from GaAs/AlAs heterostructures indicate different mechanisms of the voltage-controlled long-range IX propagation in MoSe$_2$/WSe$_2$ heterostructures. These mechanisms are discussed below.

IXs are the lowest energy exciton state in MoSe$_2$/WSe$_2$ heterostructures even at no applied voltage (Fig. 1a). The absence of long-range IX propagation at $V_g = 0$ indicates IX localization. As outlined above, the moiré superlattice potentials are predicted to cause modulations of exciton energy reaching tens of meV [31–34] and these strong energy modulations can lead to the exciton local-ization.

The regime of long-range IX propagation is realized at $V_g \gtrsim 8$ V (Fig. 1c,d). In this regime, the IX energy in the drain is effectively controlled by voltage applied to the graphene electrode (Fig. 2). This indicates metallization of the heterostructure that extends the voltage application beyond the region of the graphene electrode over the entire heterostructure. The metallization contributes to screening of in-plane potential landscapes, facilitating long-range IX propagation.

The IX propagation is enhanced at higher excitation powers (Fig. 3a-c). A similar behavior is observed for IXs in GaAs heterostructures [5]. IXs are out-of-plane dipoles, which interact repulsively. The repulsive interaction between IXs contributes to screening of in-plane potential landscapes. Furthermore, due to the repulsive interaction, the IX energy increases with density and, as a result, is higher in the region of IX generation (Fig. 3d). This results in IX drift away from the excitation region. Both enhanced screening of in-plane potentials and drift enhance the IX propagation with increasing density [5].

Increasing $V_g$ also increases the IX lifetime (Fig. S2). The increase of IX lifetime contributes to the enhancement of IX propagation with voltage. The IX diffusion coefficient $D$ can be estimated from the IX propagation length $l$ and IX lifetime $\tau$ as $D \sim l^2/\tau$. For $l \sim 10$ μm corresponding to the IX propagation length (Fig. 1) and $\tau \sim 1$ μs corresponding to the IX luminescence decay time at long delay times (Fig. S2) at $V_g = 10$ V, the estimate gives $D \sim 1$ cm$^2$/s.

This work presents the proof of principle to overcome moiré superlattice potentials and realize the long-range IX propagation in MoSe$_2$/WSe$_2$ TMD heterostructures that make TMD a promising materials platform for the development of excitonic devices. The IX binding energies are high enough to make the IXs stable at room temperature and IXs in the laser excitation spot are observed at room temperature in MoSe$_2$/WSe$_2$ heterostructures [53]. However, the long-range IX propagation through the drain region and the switching between the on and off state are observed up to $\sim 50$ K in the studied heterostructure (Fig. 4), presumably due to the heterostructure imperfections. The realization of long-range IX propagation at higher temperatures is the subject for future works.

In summary, we trace the IX luminescence along the IX propagation path and control the IX propagation by voltage in a MoSe$_2$/WSe$_2$ heterostructure. The $1/e$ IX luminescence decay distances reach 13 μm. The IX luminescence signal in the drain of the excitonic transistor is controlled within 40 times by voltage on the gate electrode. This presents the direct measurement of long-range IX propagation and excitonic transistor action in van der Waals TMD heterostructures. The data show that the long-range IX propagation is possible in TMD heterostructures with the moiré superlattice potentials.

**Acknowledgments.** We thank E.V. Calman, M.M. Fogler, S. Hu, A. Mishchenko, and A.K. Geim for valuable discussions and contributions at the earlier stage of studies of IXs in TMD heterostructures. These studies were supported by DOE Office of Basic Energy Sciences under award DE-FG02-07ER46449. The heterostructure fabrication and data analysis were supported by SRC and NSF grant 1905478.

## References

[1] Y.E. Lozovik, V.I. Yudson, A new mechanism for superconductivity: pairing between spatially separated electrons and holes, *Sov. Phys. JETP* **44**, 389 (1976).

[2] M. Hagn, A. Zrenner, G. Böhm, G. Weimann, Electric-field-induced exciton transport in coupled quantum well structures, *Appl. Phys. Lett.* **67**, 232 (1995).

[3] L.V. Butov, A.C. Gossard, D.S. Chemla, Macroscopically ordered state in an exciton system, *Nature* **418**, 751 (2002).

[4] Z. Vörös, R. Balili, D.W. Snoke, L. Pfeiffer, K. West, Long-Distance Diffusion of Excitons in Double Quantum Well Structures, *Phys. Rev. Lett.* **94**, 226401 (2005).

[5] A.L. Ivanov, L.E. Smallwood, A.T. Hammack, Sen Yang, L.V. Butov, A.C. Gossard, Origin of the inner ring in photoluminescence patterns of quantum well excitons, *Europhys. Lett.* **73**, 920 (2006).

[6] A. Gärtner, A.W. Holleitner, J.P. Kotthaus, D. Schuh, Drift mobility of long-living excitons in coupled GaAs quantum wells, *Appl. Phys. Lett.* **89**, 052108 (2006).

[7] S. Lazić, P.V. Santos, R. Hey, Exciton transport by moving strain dots in GaAs quantum wells, *Phys. E* **42**, 2640 (2010).

[8] A.V. Gorbunov, V.B. Timofeev, D.A. Demin, Electro-Optical Trap for Dipolar Excitons in a GaAs/AlAs Schottky Diode with a Single Quantum Well, *JETP Letters* **94**, 800 (2011).

[9] M. Alloing, A. Lemaître, E. Galopin, F. Dubin, Nonlinear dynamics and inner-ring photoluminescence pattern of indirect excitons, *Phys. Rev. B* **85**, 245106 (2012).

[10] S. Lazić, A. Violante, K. Cohen, R. Hey, R. Rapaport, P.V. Santos, Scalable interconnections for remote indirect exciton systems based on acoustic transport, *Phys. Rev. B* **89**, 085313 (2014).

[11] C.J. Dorow, M.W. Hasling, D.J. Choksy, J.R. Leonard, L.V. Butov, K.W. West, L.N. Pfeiffer, High-mobility indirect excitons in wide single quantum well, *Appl. Phys. Lett.* **113**, 212102 (2018).

[12] S. Zimmermann, G. Schedelbeck, A.O. Govorov, A. Wixforth, J.P. Kotthaus, M. Bichler, W. Wegscheider, G. Abstreiter, Spatially resolved exciton trapping in a voltage-controlled lateral superlattice, *Appl. Phys. Lett.* **73**, 154 (1998).

[13] M. Remeika, J.C. Graves, A.T. Hammack, A.D. Meyertholen, M.M. Fogler, L.V. Butov, M. Hanson and A.C. Gossard, Localization-Delocalization Transition of Indirect Excitons in Lateral Electrostatic Lattices, *Phys. Rev. Lett.* **102**, 186803 (2009).

[14] M. Remeika, M.M. Fogler, L.V. Butov, M. Hanson, A.C. Gossard, Two-Dimensional Electrostatic Lattices for Indirect Excitons, *Appl. Phys. Lett.* **100**, 061103 (2012).

[15] X.P. Vögele, D. Schuh, W. Wegscheider, J.P. Kotthaus, A.W. Holleitner, Density Enhanced Diffusion of Dipolar Excitons within a One-Dimensional Channel, *Phys. Rev. Lett.* **103**, 126402 (2009).

[16] K. Cohen, R. Rapaport, P.V. Santos, Remote Dipolar Interactions for Objective Density Calibration and Flow Control of Excitonic Fluids, *Phys. Rev. Lett.* **106**, 126402 (2011).

[17] A.G. Winbow, J.R. Leonard, M. Remeika, Y.Y. Kuznetsova, A.A. High, A.T. Hammack, L.V. Butov, J. Wilkes, A.A. Guenther, A.L. Ivanov, M. Hanson, A.C. Gossard, Electrostatic Conveyer for Excitons, *Phys. Rev. Lett.* **106**, 196806 (2011).

[18] C.J. Dorow, J.R. Leonard, M.M. Fogler, L.V. Butov, K.W. West, L.N. Pfeiffer, Split-gate device for indirect excitons, *Appl. Phys. Lett.* **112**, 183501 (2018).

[19] A.A. High, E.E. Novitskaya, L.V. Butov, M. Hanson, A.C. Gossard, Control of exciton fluxes in an excitonic integrated circuit, *Science* **321**, 229 (2008).

[20] G. Grosso, J. Graves, A.T. Hammack, A.A. High, L.V. Butov, M. Hanson, A.C. Gossard, Excitonic switches operating at around 100 K, *Nature Photonics* **3**, 577-580 (2009).

[21] D.S. Chemla, D.A.B. Miller, P.W. Smith, A.C. Gossard, W. Wiegmann, Room temperature excitonic nonlinear absorption and refraction in GaAs/AlGaAs multiple quantum well structures, *IEEE J. Quantum Electron.* **20**, 265-275 (1984).

[22] K. Sivalertporn, L. Mouchliadis, A.L. Ivanov, R. Philp, E.A. Muljarov, Direct and indirect excitons in semiconductor coupled quantum wells in an applied electric field, *Phys. Rev. B* **85**, 045207 (2012).

[23] A. Zrenner, P. Leeb, J. Schäfler, G. Böhm, G. Weimann, J.M. Worlock, L.T. Florez, J.P. Harbison, Indirect excitons in coupled quantum well structures, *Surf. Sci.* **263**, 496-501 (1992).

[24] François Chiaruttini, Thierry Guillet, Christelle Brimont, Benoit Jouault, Pierre Lefebvre, Jessica Vives, Sebastien Chenot, Yvon Cordier, Benjamin Damilano, Maria Vladimirova, Trapping Dipolar Exciton Fluids in GaN/(AlGa)N Nanostructures, *Nano Lett.* **19**, 4911-4918 (2019).

[25] A.K. Geim, I.V. Grigorieva, Van der Waals heterostructures, *Nature* **499**, 419-425 (2013).

[26] Ziliang Ye, Ting Cao, Kevin O'Brien, Hanyu Zhu, Xiaobo Yin, Yuan Wang, Steven G. Louie, Xiang Zhang, Probing excitonic dark states in single-layer tungsten disulphide, *Nature* **513**, 214-218 (2014).

[27] Alexey Chernikov, Timothy C. Berkelbach, Heather M. Hill, Albert Rigosi, Yilei Li, Ozgur Burak Aslan, David R. Reichman, Mark S. Hybertsen, Tony F. Heinz, Exciton Binding Energy and Nonhydrogenic Rydberg Series in Monolayer $WS_2$, *Phys. Rev. Lett.* **113**, 076802 (2014).

[28] M.M. Fogler, L.V. Butov, K.S. Novoselov, High-temperature superfluidity with indirect excitons in van der Waals heterostructures, *Nature Commun.* **5**, 4555 (2014).

[29] Thorsten Deilmann, Kristian Sommer Thygesen, Interlayer Trions in the $MoS_2/WS_2$ van der Waals Heterostructure, *Nano Lett.* **18**, 1460 (2018).

[30] E.V. Calman, M.M. Fogler, L.V. Butov, S. Hu, A. Mishchenko, A.K. Geim, Indirect excitons in van der Waals heterostructures at room temperature, *Nature Commun.* **9**, 1895 (2018).

[31] Fengcheng Wu, Timothy Lovorn, A.H. MacDonald, Theory of optical absorption by interlayer excitons in transition metal dichalcogenide heterobilayers, *Phys. Rev. B* **97**, 035306 (2018).

[32] Hongyi Yu, Gui-Bin Liu, Wang Yao, Brightened spin-triplet interlayer excitons and optical selection rules in van der Waals heterobilayers, *2D Mater.* **5**, 035021 (2018).

[33] Fengcheng Wu, Timothy Lovorn, A.H. MacDonald, Topological Exciton Bands in Moiré Heterojunctions, *Phys. Rev. Lett.* **118**, 147401 (2017).

[34] Hongyi Yu, Gui-Bin Liu, Jianju Tang, Xiaodong Xu, Wang Yao, Moiré excitons: From programmable quantum emitter arrays to spin-orbit–coupled artificial lattices, *Sci. Adv.* **3**, e1701696 (2017).

[35] Nardeep Kumar, Qiannan Cui, Frank Ceballos, Dawei He, Yongsheng Wang, Hui Zhao, Exciton diffusion in monolayer and bulk $MoSe_2$, *Nanoscale* **6**, 4915, (2014).

[36] Marvin Kulig, Jonas Zipfel, Philipp Nagler, Sofia Blanter, Christian Schüller, Tobias Korn, Nicola Paradiso, Mikhail M. Glazov, Alexey Chernikov, Exciton Diffusion and Halo Effects in Monolayer Semiconductors, *Phys. Rev. Lett.* **120**, 207401 (2018).

[37] F. Cadiz, C. Robert, E. Courtade, M. Manca, L. Martinelli, T. Taniguchi, K. Watanabe, T. Amand, A.C.H. Rowe, D. Paget, B. Urbaszek, X. Marie, Exciton diffusion in $WSe_2$ monolayers embedded in a van der Waals heterostructure, *Appl. Phys. Lett.* **112**, 152106 (2018).

[38] Darwin F. Cordovilla Leon, Zidong Li, Sung Woon Jang, Che-Hsuan Cheng, Parag B. Deotare, Exciton transport in strained monolayer $WSe_2$, *Appl. Phys. Lett.* **113**, 252101 (2018).

[39] Darwin F. Cordovilla Leon, Zidong Li, Sung Woon Jang, Parag B. Deotare, Hot exciton transport in $WSe_2$ monolayers, *Phys. Rev. B* **100**, 241401(R) (2019).

[40] Shengcai Hao, Matthew Z. Bellus, Dawei He, Yongsheng Wang, Hui Zhao, Controlling exciton transport in monolayer $MoSe_2$ by dielectric screening, *Nanoscale Horiz.* **5**, 139 (2020).

[41] Pasqual Rivera, Kyle L. Seyler, Hongyi Yu, John R. Schaibley, Jiaqiang Yan, David G. Mandrus, Wang Yao, Xiaodong Xu, Valley-polarized exciton dynamics in a 2D semiconductor heterostructure, *Science* **351**, 688-690 (2016).

[42] Luis A. Jauregui, Andrew Y. Joe, Kateryna Pistunova, Dominik S. Wild, Alexander A. High, You Zhou, Giovanni Scuri, Kristiaan De Greve, Andrey Sushko, Che-Hang Yu, Takashi Taniguchi, Kenji Watanabe, Daniel J. Needleman, Mikhail D. Lukin, Hongkun Park, Philip Kim, Electrical control of interlayer exciton dynamics in atomically thin heterostructures, *Science* **366**, 870-875 (2019).

[43] Dmitrii Unuchek, Alberto Ciarrocchi, Ahmet Avsar, Zhe Sun, Kenji Watanabe, Takashi Taniguchi, Andras Kis, Valley-polarized exciton currents in a van der Waals heterostructure, *Nature Nanotechnology* **14**, 1104-1109 (2019).

[44] Yuanda Liu, Kostya S. Novoselov, Weibo Gao, Electrically controllable router of interlayer excitons, arXiv:1911.12061 (2019).

[45] Zumeng Huang, Yuanda Liu, Kevin Dini, Qinghai Tan, Zhuojun Liu, Hanlin Fang, Jin Liu, Timothy Liew, Weibo Gao, Robust room temperature valley Hall effect of interlayer excitons, *Nano Lett.* **20**, 1345-1351 (2020).

[46] Dmitrii Unuchek, Alberto Ciarrocchi, Ahmet Avsar, Kenji Watanabe, Takashi Taniguchi, Andras Kis, Room-temperature electrical control of exciton flux in a van der Waals heterostructure, *Nature* **560**, 340-344 (2019).

[47] Gui-Bin Liu, Wen-Yu Shan, Yugui Yao, Wang Yao, Di Xiao, Three-band tight-binding model for monolayers of group-VIB transition metal dichalcogenides, *Phys. Rev. B* **88**, 085433 (2013).

[48] Gang Wang, Cedric Robert, Aslihan Suslu, Bin Chen, Sijie Yang, Sarah Alamdari, Iann C. Gerber, Thierry Amand, Xavier Marie, Sefaattin Tongay, Bernhard Urbaszek, Spin-orbit engineering in transition metal dichalcogenide alloy monolayers, *Nature Commun.* **6**, 10110 (2015).

[49] Xiao-Xiao Zhang, Yumeng You, Shu Yang Frank Zhao, Tony F. Heinz, Experimental Evidence for Dark Excitons in Monolayer $WSe_2$, *Phys. Rev. Lett.* **115**, 257403 (2015).

[50] You Zhou, Giovanni Scuri, Dominik S. Wild, Alexander A. High, Alan Dibos, Luis A. Jauregui, Chi Shu, Kristiaan De Greve, Kateryna Pistunova, Andrew Y. Joe, Takashi Taniguchi, Kenji Watanabe, Philip Kim, Mikhail D. Lukin, Hongkun Park, Probing dark excitons in atomically thin semiconductors via near-field coupling to surface plasmon polaritons, *Nature Nano.* **12**, 856 (2017).

[51] Xiao-Xiao Zhang, Ting Cao, Zhengguang Lu, Yu-Chuan Lin, Fan Zhang, Ying Wang, Zhiqiang Li, James C. Hone, Joshua A. Robinson, Dmitry Smirnov, Steven G. Louie, Tony F. Heinz, Magnetic brightening and control of dark excitons in monolayer $WSe_2$, *Nature Nano.* **12**, 883 (2017).

[52] R. Bistritzer, A.H. MacDonald, Moiré butterflies in twisted bilayer graphene, *Phys. Rev. B* **84**, 035440 (2011).

[53] E.V. Calman, L.H. Fowler-Gerace, D.J. Choksy, L.V. Butov, D.E. Nikonov, I.A. Young, S. Hu, A. Mishchenko, A.K. Geim, Indirect Excitons and Trions in $MoSe_2/WSe_2$ van der Waals Heterostructures, *Nano Lett.* **20**, 1869 (2020).

# Supplementary information: Voltage-controlled long-range propagation of indirect excitons in van der Waals heterostructure

L. H. Fowler-Gerace,[1] D. J. Choksy,[1] and L. V. Butov[1]

[1]Department of Physics, University of California at San Diego, La Jolla, CA 92093, USA

(Dated: January 4, 2021)

PACS numbers:

### IX propagation for different excitation spot positions

The voltage-controlled long-range IX propagation is observed for the laser excitation spot positioned in the source, gate, or drain regions (Fig. S1). In all these three cases, switching on the exciton propagation by applied voltage extends IXs over the entire MoSe$_2$/WSe$_2$ heterostructure and enhances the IX luminescence $1/e$ decay distance beyond 10 $\mu$m (Fig. S1).

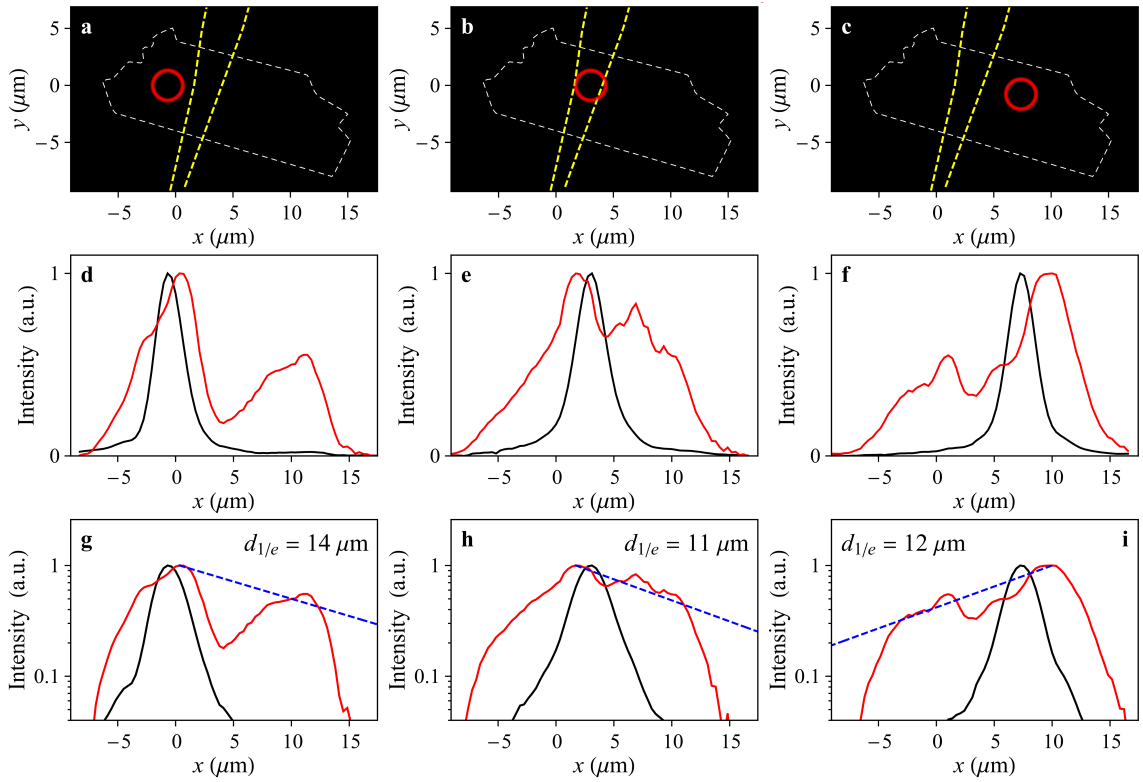

FIG. S1: **Voltage-controlled IX propagation.** (a-c) The white and yellow dashed lines show the boundaries of MoSe$_2$/WSe$_2$ heterostructure and graphene gate, respectively. The laser excitation spot [red circle in (a-c)] is positioned at the source (a,d,g), gate (b,e,h), and drain (c,f,i) region of the device. The gate voltage $V_g$ controls the IX propagation from the laser excitation spot. (d-f) Normalized IX luminescence profiles along $y = 0$ for $V_g = 10$ V (red lines) and 0 (black lines). (g-i) Same IX luminescence profiles on log scale. For comparison with the IX luminescence profiles at $V_g = 10$ V, dashed lines show exponential signal reduction with $1/e$ decay distance $d_{1/e} = 14$, 11, and 12 $\mu$m, respectively. A lower IX luminescence intensity is seen in the region covered by the graphene gate, which is centered at $x = 4$ $\mu$m. For all data, $P_{ex} = 4$ mW, $T = 1.7$ K.

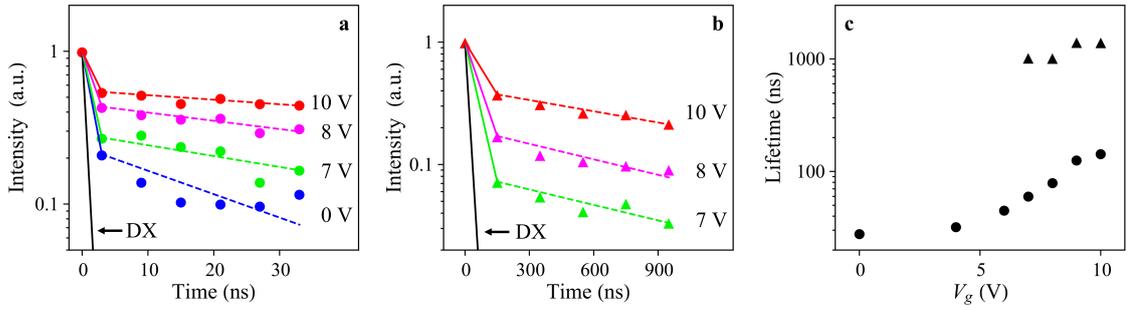

FIG. S2: **IX luminescence decay.** (a,b) Decay of IX luminescence measured at the IX line energies 1.24-1.38 eV at short (a) and long (b) delay times vs. voltage. This energy range corresponds to the IX state showing the long-range propagation (Fig. 2). For the data in (a), the laser excitation pulse duration $\tau_{pulse}$ = 14 ns, period $\tau_{period}$ = 60 ns, and edge sharpness ∼ 0.5 ns and the signal integration window $\tau_w$ = 6 ns. For the data in (b), to access the long delay times $\tau_{pulse}$, $\tau_{period}$, and $\tau_w$ are increased to 400, 1600, and 200 ns, respectively. (c) IX luminescence decay times at short (points) and long (triangles) delay times derived from the data (a) and (b), respectively. The IX luminescence is integrated over the entire heterostructure to increase the signal. For all data, $P_{ex}$ = 4 mW, $T$ = 1.7 K.

### IX luminescence decay

The IX luminescence decay is non-exponential. We probe the faster initial decay (Fig. S2a) with the laser excitation pulse duration $\tau_{pulse}$ = 14 ns, period $\tau_{period}$ = 60 ns, and edge sharpness ∼ 0.5 ns and the signal integration window $\tau_w$ = 6 ns. We probe a slower decay at long delay times with $\tau_{pulse}$, $\tau_{period}$, and $\tau_w$ increased to 400, 1600, and 200 ns, respectively. The IX luminescence decay times increase with $V_g$ (Fig. S2c). The DX decay closely follows the excitation laser decay indicating that the DX lifetime is shorter than the experimental resolution.

The IX luminescence decay time increase with $V_g$ is in accord with the reduction of IX energy (Fig. 2b) further below the DX energy that reduces the overlap of the electron and hole wavefunctions for IXs [1]. The IX decay times are orders of magnitude longer than the DX decay times [2] and are controlled by gate voltage (Fig. S2). Different factors may contribute to deviations of the luminescence decay from an exponential decay. For instance, due to a possible heterostructure inhomogeneity the areas with shorter exciton lifetimes may contribute more at initial decay times. A fast initial component may also appear due to the decay of low-energy DX states, which appear in the IX spectral range due to the tail of DX density of states. Localized DXs at low energies in the spectral range of IXs were studied in GaAs/AlAs heterostructure [3].

### Optical measurements

In the cw experiments, excitons were generated by a cw HeNe laser with excitation energy $E_{ex}$ = 1.96 eV. Luminescence spectra were measured using a spectrometer with resolution 0.2 meV and a liquid-nitrogen-cooled CCD. The laser was focused to a spot size ∼ 3.5 µm. The IX luminescence decay time was measured using a pulsed semiconductor laser with $E_{ex}$ = 1.96 eV, the emitted light was detected by a liquid-nitrogen-cooled CCD coupled to a PicoStar HR TauTec time-gated intensifier. The experiments were performed in a variable-temperature 4He cryostat.

### The heterostructure fabrication and characterization

The studied van der Waals heterostructures were assembled using the peel-and-lift technique [4]. In brief, individual crystals of graphene, hBN, MoSe$_2$, and WSe$_2$ were first micromechanically exfoliated onto different Si substrates that were coated with a double polymer layer consisting of polymethyl glutarimide (PMGI) and polymethyl methacrylate (PMMA). The bottom PMGI was then dissolved releasing the top PMMA membrane with a target 2D crystal. Separately, a large graphite crystal was exfoliated onto an oxidized Si wafer, which later served as the bottom electrode. The PMMA membrane was then flipped over and aligned above an atomically-flat region of the graphite crystal using a micromechanical transfer stage. The two crystals were brought into contact and the temperature of the stage was ramped to 80º C in order to increase adhesion between the 2D crystals. Then, the PMMA membrane was slowly peeled off leaving the bilayer stack on the wafer. The procedure was repeated leading to a multi-crystal stack with the desired layer sequence. No intentional sample doping was done, however, unintentional *n*-type doping is typical for TMD layers [4]. The thickness of bottom and top hBN layers is about 40 and 30 nm, respectively. The MoSe$_2$ layer is on top of the WSe$_2$ layer. The long edges of MoSe$_2$ and WSe$_2$ monolayers were aligned. The long WSe$_2$ and MoSe$_2$ edges reach ∼ 30 and ∼ 20 $\mu$m, respectively, that enables a rotational alignment between WSe$_2$ and MoSe$_2$ monolayers within $2^0$ (Figs. 1b and S3). Figure S3 presents a microscope image showing the layer pattern of the device on an area larger than in Fig. 1b. The layer boundaries are indicated. The hBN layers cover the entire areas of MoSe$_2$, WSe$_2$, and graphene layers.

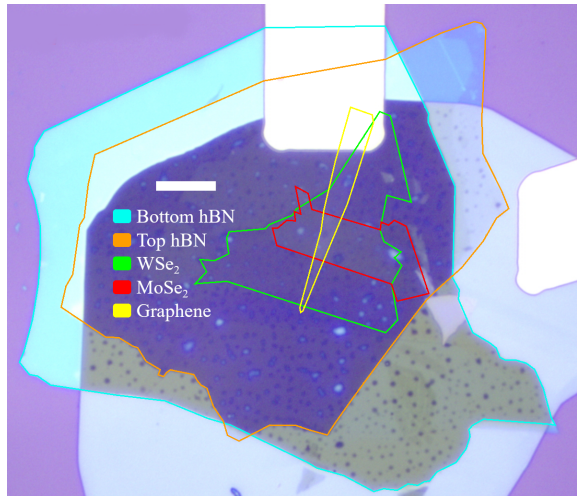

FIG. S3: **The layer pattern.** Microscope image showing the layer pattern of the device on an area larger than in Fig. 1b. Scale bar is 10 $\mu$m. The green, red, yellow, cyan, and orange lines indicate the boundaries of WSe$_2$ and MoSe$_2$ monolayers, graphene gate, and bottom and top hBN layers, respectively.

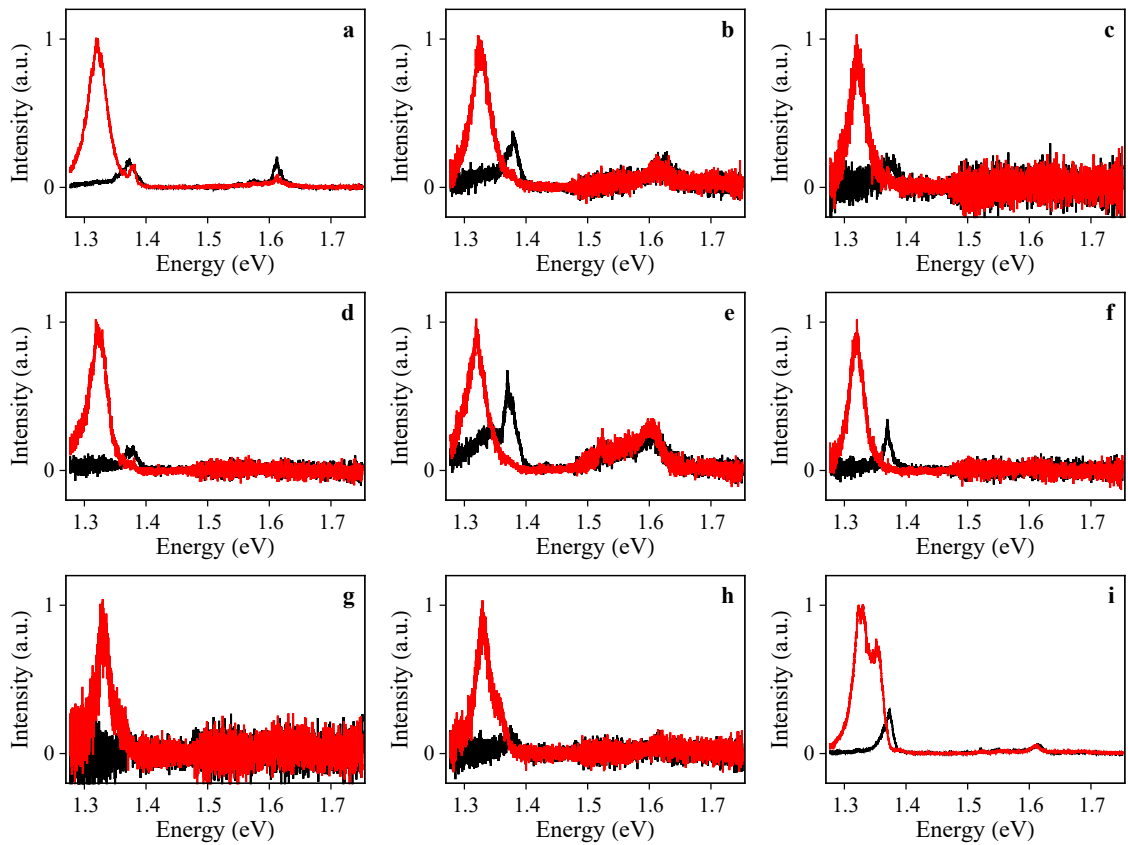

FIG. S4: **Luminescence spectra for an extended spectral range.** Luminescence spectra for an extended spectral range including the DX luminescence for excitation in the source (a-c), gate (d-f), and drain (g-i) at $x = 1, 4$, and $9$ $\mu$m, respectively, measured in the source (a, d, g), gate (b, e, h), and drain (c, f, i) at $x = 0, 4$, and $9$ $\mu$m, respectively, for gate voltages $V_g = 0$ and 10 V corresponding to the off (black) and on (red) state of the excitonic transistor. $P_{ex} = 0.05$ mW, $T = 1.7$ K.

Photolithography was employed to define contact regions to graphene and graphite crystals, which followed by metal evaporation of electrical contacts (10 nm Ti / 300 nm Au). Lift-off and sample annealing were performed by immersing the lithographically patterned sample in Remover PG, an N-Methyl-2-pyrrolidone (NMP) based solvent stripper, at 70º C for 12 hours.

The voltage was applied to the graphene top gate, the graphite back gate was grounded. The leakage current across the device in the on state reached ∼ 0.7 nA at the highest voltage (15 V) and excitation power (4 mW) in the experiment.

Luminescence spectra for an extended spectral range including the DX luminescence are presented in Fig. S4. Figure S4 shows the spectra under the graphene gate, on the source, and drain parts of the structure at gate voltages corresponding to the on and off states of the excitonic transistor.

**References**

[1] A. Zrenner, P. Leeb, J. Schäfler, G. Böhm, G. Weimann, J.M. Worlock, L.T. Florez, J.P. Harbison, Indirect excitons in coupled quantum well structures, *Surf. Sci.* **263**, 496-501 (1992).
[2] T. Korn, S. Heydrich, M. Hirmer, J. Schmutzler, C. Schüller, Low-temperature photocarrier dynamics in monolayer MoS$_2$, *Appl. Phys. Lett.* **99**, 102109 (2011).
[3] A. Zrenner, L.V. Butov, M. Hagn, G. Abstreiter, G. Böhm, G. Weimann, Quantum dots formed by interface fluctuations in AlAs/GaAs coupled quantum well structures, *Phys. Rev. Lett.* **72**, 3382-3385 (1994).
[4] F. Withers, O. Del Pozo-Zamudio, A. Mishchenko, A.P. Rooney, A. Gholinia, K. Watanabe, T. Taniguchi, S.J. Haigh, A.K. Geim, A.I. Tartakovskii, K.S. Novoselov. Light-emitting diodes by band-structure engineering in van der Waals heterostructures, *Nature Materials* **14**, 301-306 (2015).